\documentclass[conference, 10pt]{IEEEtran}

\usepackage{psfrag}
\usepackage{epsfig}
\usepackage{amsmath}
\usepackage{amssymb}
\usepackage{amsfonts}
\usepackage{color}
\usepackage{algorithm,algorithmic}
\usepackage{epstopdf}
\usepackage{psfrag}

\def\punit{\, \mathrm}

\def\e{\mathrm{ e}}
\newcommand{\argmax}{\mathop{\mathrm{argmax}}}

\newcommand{\algorithmicinput}{\textbf{input:}}
\newcommand{\INPUT}{\item[\algorithmicinput]}
\newcommand{\algorithmicoutput}{\textbf{output:}}
\newcommand{\OUTPUT}{\item[\algorithmicoutput]}

\begin{document}

	
\title{Optimized Processing Order for 3D Hole Filling in Video Sequences Using Frequency Selective Extrapolation}

\author{\IEEEauthorblockN{J{\"u}rgen~Seiler, Susanne Sch{\"o}ll, Wolfgang Schnurrer, and~Andr{\'e}~Kaup}
	\IEEEauthorblockA{Chair of Multimedia Communications and Signal Processing, Friedrich-Alexander-Universit{\"a}t Erlangen-N{\"u}rnberg (FAU)\\ Cauerstr. 7, 91058 Erlangen, Germany\\
		juergen.seiler@FAU.de; susanne.schoell@FAU.de; wolfgang.schnurrer@FAU.de; \mbox{andre.kaup@FAU.de}\\}}

\maketitle

\fontsize{9.3}{10.1}\selectfont


\begin{abstract} \label{abstract}
A problem often arising in video communication is the reconstruction of missing or distorted areas in a video sequence. Such holes of unavailable pixels may be caused for example by transmission errors of coded video data or undesired objects like logos. In order to close the holes given neighboring available content, a signal extrapolation has to be performed. The best quality can be achieved, if spatial as well as temporal information is used for the reconstruction. However, the question always is in which order to process the extrapolation to obtain the best result. In this paper, an optimized processing order is introduced for improving the extrapolation quality of Three-dimensional Frequency Selective Extrapolation. Using the proposed optimized order, holes in video sequences can be closed from the outer margin to the center, leading to a higher reconstruction quality, and visually noticeable gains of more than 0.5 dB PSNR are possible. 
\end{abstract}


\section{Introduction} \label{sec:introduction}
The extrapolation of unknown signal content is a crucial task in video communication. Even though in the ideal case, all pixels in a video sequence would be known, there exist many scenarios where a video sequence is not completely known, but rather contains holes where the information is missing. In video communication this happens for example if errors occur during the transmission of coded video data. In this case, the sequence cannot be decoded correctly and contains regions of unknown content. In order to provide a decent presentation quality and to reduce error propagation, error concealment can be applied for reconstructing the unknown content \cite{Wang1998, Koloda2013, Zhang2000a}. Another application for signal extrapolation is the removal of logos from video signals \cite{Kuo2008, Dashti2015}. Aside from this, from a more general point of view, the reconstruction of missing or distorted regions in a video sequence can be regarded as inpainting problem \cite{Chang2008, Shih2009, Ebdelli2012, Newson2014} or a gap filling problem \cite{Garcia2010, Wang2012}. 

\begin{figure}
	\psfrag{x}[c][c][0.75]{$x$}
	\psfrag{y}[c][c][0.75]{$y$}
	\psfrag{t}[c][c][0.75]{$t$}
	\includegraphics[width=0.46\textwidth]{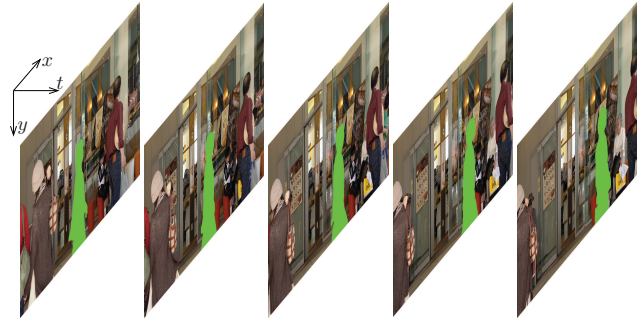}
	\caption{Extrapolation example: Removing a person (marked in green) from a video sequence $v\left[x,y,t\right]$ causes an arbitrarily shaped hole to be closed. }
	\label{fig:sequence}
\end{figure}

Independent of the actual application, the extrapolation scenario can be regarded as follows. The video sequence forms a three-dimensional volume and this volume contains arbitrarily shaped three-dimensional holes as shown in Fig.\ \ref{fig:sequence} for an example. In order to reconstruct the video sequence, these holes have to be filled, or respectively, extrapolated, given the available samples. The algorithms to achieve this can be divided into three groups. First, a two-dimensional extrapolation within every frame of the sequence can be applied, making use of concepts as shown for example in \cite{Guillemot2014}. For this, no temporal information is used. Second, a temporal extrapolation can be used as shown for example in \cite{Zhang2000a}. There, the missing area in one frame is reconstructed by inserting a fitting region from a neighboring completely known frame. Even though these concepts can achieve a decent quality in some scenarios, they only perform inferior if confronted with arbitrarily shaped holes as they neglect either the spatial or the temporal correlations within a video sequence. Thus, the third group of algorithms is better suited, which is spatio-temporal extrapolation that exploits temporal and spatial information at the same time. By exploiting most of the available information for the reconstruction, spatio-temporal extrapolation typically achieves the highest reconstruction quality \cite{Tsekeridou2000}.

In \cite{Meisinger2007}, Three-dimensional Frequency Selective Extrapolation (3D-FSE) was introduced as a powerful spatio-temporal signal extrapolation algorithm. Even though 3D-FSE can achieve a high reconstruction quality, it is not suited well for filling arbitrarily shaped holes in video sequences up to now. The reason for this is that the processing order is not adaptive to the shape of the area to be extrapolated. 

This paper introduces a novel optimized three-dimensional processing order for 3D-FSE which is able to effectively close arbitrarily shaped holes in video sequences. At the same time, the proposed processing order allows for a parallel computation which has not been possible, so far. In the next section, 3D-FSE is briefly revisited. Furthermore, some extensions originally developed for a two-dimensional extrapolation are transferred to 3D-FSE and it is generalized in order to be able to work on arbitrarily shaped holes. Afterwards, the new optimized processing order is introduced, before simulations are provided for showing its effectiveness.


\section{Generalized Three-Dimensional Frequency Selective Extrapolation} \label{sec:generalized_3D-FSE} 

The Three-dimensional Frequency Selective Extrapolation (3D-FSE) as it is proposed in \cite{Meisinger2007} is not well suited for closing arbitrarily shaped three-dimensional holes in video sequences. The reason for this is that the original 3D-FSE is only designed for extrapolating block losses or connected block losses. Furthermore, the frequency-domain implementation proposed in \cite{Meisinger2007} generates a real-valued model which causes an unnecessary high computational complexity. In contrast to this, two-dimensional Frequency Selective Extrapolation (2D-FSE) has gained a lot of enhancements in the past years as, e.g., orthogonality deficiency compensation \cite{Seiler2007, Seiler2008} or the more efficient complex-valued model generation \cite{Seiler2010c}. Hence, in the following, a generalized 3D-FSE is outlined, including the available enhancements from the two-dimensional case and therewith enabling it for high quality extrapolation of arbitrarily shaped holes.

\begin{figure}
	\centering
	\psfrag{m}[c][c][0.75]{$m$}
	\psfrag{n}[c][c][0.75]{$n$}
	\psfrag{p}[c][c][0.75]{$p$}
	\psfrag{A}[c][c]{$\mathcal{A}$}
	\psfrag{Bi}[c][c]{\color{white}$\mathcal{B}_i$}
	\psfrag{Bo}[c][c]{\color{white}$\mathcal{B}_o$}
	\psfrag{R}[c][c]{$\mathcal{R}$}
	\psfrag{c}[c][c]{$\mathfrak{c}$}
	\includegraphics[width=0.4\textwidth]{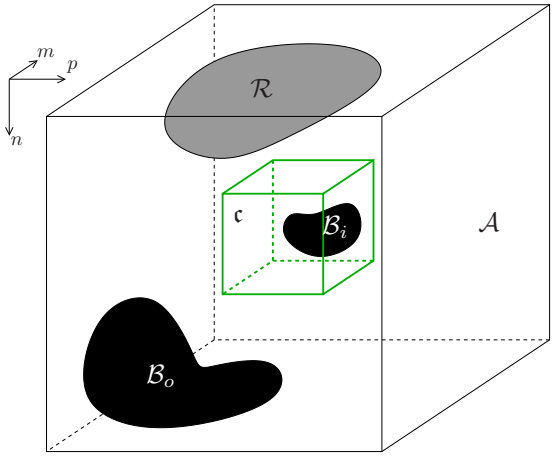}
	\caption{3D extrapolation volume $\mathcal{L}$ for reconstruction of cube $\mathfrak{c}$, located in the center. $\mathcal{L}$ consists of original samples subsumed in $\mathcal{A}$, already reconstructed samples subsumed in $\mathcal{R}$, unknown samples outside cube $\mathfrak{c}$, subsumed in $\mathcal{B}_{o}$, and inside cube $\mathfrak{c}$, subsumed in $\mathcal{B}_{i}$.}
	\label{fig:cube_3D}
\end{figure}

For the extrapolation, we regard a video sequence $v\left[x,y,t\right]$ with spatial coordinates $x$ and $y$ and temporal coordinate $t$. The video sequence contains arbitrarily shaped holes which have to be filled by extrapolation as shown in Fig.\ \ref{fig:sequence}. In order to close the holes, the volume defined by the video sequence is split into equally-sized cubes and all the cubes containing holes actually have to be processed. For extrapolating the unknown samples within a cube $\mathfrak{c}$, the cube itself together with a neighborhood of samples belonging to adjacent cubes is considered as shown in Fig.\ \ref{fig:cube_3D}. Cube $\mathfrak{c}$ together with the considered neighborhood forms extrapolation volume $\mathcal{L}$ with the three coordinates $m$, $n$, and $p$ and is of size $M$, $N$, $P$. The samples inside volume $\mathcal{L}$ can be divided into four groups. All originally known and therefore available samples are subsumed in $\mathcal{A}$, samples belonging to neighboring cubes that have been reconstructed before belong to $\mathcal{R}$, all unknown samples inside the cube $\mathfrak{c}$ are subsumed in $\mathcal{B}_i$ and outside $\mathfrak{c}$ are subsumed in $\mathcal{B}_o$.

The objective of 3D-FSE is to generate a three-dimensional model $g\left[m,n,p\right]$ for the signal $s\left[m,n,p\right]$ in volume $\mathcal{L}$ as weighted superposition of the Fourier basis functions
\begin{equation}
\label{eq:fourier_bf}
\varphi_{\left(k,l,q\right)}\left[m,n,p\right] = \e^{\j \frac{2\pi}{M}km}\e^{\j \frac{2\pi}{N}ln}\e^{\j \frac{2\pi}{P}qp}.
\end{equation}
In the original 3D-FSE \cite{Meisinger2007}, the objective was to generate the real-valued model
\begin{equation}
\label{eq:rFSE_model}
g\left[m,n,p\right] \hspace{-1mm}=\hspace{-5mm} \sum_{\left(k,l,q\right)\in\mathcal{K}}\hspace{-4mm} \frac{\hat{c}_{\left(k,l,q\right)} \varphi_{\left(k,l,q \right)}^\ast\hspace{-1mm}\left[m,n,p\right] + \hat{c}_{\left(k,l,q \right)}^\ast \varphi_{\left(k,l,q \right)}\hspace{-1mm}\left[m,n,p\right]}{2}
\end{equation}
with expansion coefficients $\hat{c}_{\left(k,l,q \right)}$ and set $\mathcal{K}$ containing the used basis functions. However, as shown in \cite{Seiler2010c}, generating a real-valued model is not necessary and only increases the computational complexity significantly. Instead, for the generalized \mbox{3D-FSE}, it is proposed to generate the complex-valued model
\begin{equation}
\label{eq:cFSE_model}
g\left[m,n,p\right] = \sum_{\left(k,l,q\right)\in\mathcal{K}}\hat{c}_{\left(k,l,q \right)} \varphi_{\left(k,l,q \right)}\left[m,n,p\right] .
\end{equation}

As shown in \cite{Seiler2010c} for the two-dimensional case, the model generation works iteratively. In doing so, in every iteration one basis function is selected and added to the model, appropriately weighted. For this, at the beginning of every iteration $\nu$, the approximation residual 
\begin{equation}
r^{\left(\nu-1\right)} \left[m,n,p\right] = s\left[m,n,p\right] - g^{\left(\nu-1\right)} \left[m,n,p\right]
\end{equation}
with respect to the previous iteration is calculated. Initially, the model $g^{\left(0\right)} \left[m,n,p\right]$ is set to zero. Next, a weighted projection of the residual onto all basis functions is performed. This yields the projection coefficients
\begin{equation}
p_{\left(k,l,q \right)}^{\left(\nu\right)} \hspace{-1mm}=\hspace{-1mm} \frac{\displaystyle \hspace{-2mm}\sum_{\left(m,n,p\right)\in\mathcal{L}}\hspace{-4mm} r^{\left(\nu-1\right)}\hspace{-1mm} \left[m,n,p\right] \varphi^\ast_{\left(k,l,q \right)}\hspace{-1mm}\left[m,n,p\right] w\left[m,n,p\right]}{\displaystyle \hspace{-2mm}\sum_{\left(m,n,p\right)\in\mathcal{L}}\hspace{-4mm} \left|\varphi_{\left(k,l,q \right)}\left[m,n,p\right]\right|^2 w\left[m,n,p\right]}
\end{equation}
for all basis function indices $\left(k,l,q \right)$. In this context, the weighting function $w\left[m,n,p\right]$ is applied to assign different weights to the samples and therewith different influence on the model generation, depending on their position. The weighting function is defined by 
\begin{equation}
\label{eq:weighting_function}
w\left[m,n,p\right] = \left\{ \begin{array}{ll} \rho\left[m,n,p\right] & \mbox{for } \left(m,n,p\right)\in \mathcal{A} \\ \delta \rho\left[m,n,p\right] & \mbox{for } \left(m,n,p\right)\in \mathcal{R} \\0 & \mbox{for } \left(m,n,p\right)\in \mathcal{B}_i \cup \mathcal{B}_o\end{array}\right. ,
\end{equation}
using the exponentially decreasing function
\begin{equation}
\rho\left[m,n,p\right] = \hat{\rho}^{\sqrt{\left(m-\frac{M-1}{2}\right)^2+\left(n-\frac{N-1}{2}\right)^2+\left(p-\frac{P-1}{2}\right)^2}}
\end{equation}
proposed in \cite{Meisinger2007}. The parameter $\hat{\rho}$ controls the speed of the decay. Using this weighting function has the following advantages. First, all unknown samples in $\mathcal{B}_i$ and $\mathcal{B}_o$ are excluded from the model generation. Second, the influence of the known samples decreases with increasing distance to cube $\mathfrak{c}$. And third, since already reconstructed samples cannot be as reliable as originally known ones, their influence can additionally be reduced by the multiplicative factor $\delta$ from the range between zero and one.

Based on the projection coefficients $p_{\left(k,l,q \right)}^{\left(\nu\right)}$, the index $\left(u,v,z\right)$ of the basis function to be added to the model can be determined by
\[
\left(u,v,z\right) =\hspace{7cm}
\]
\begin{equation}
 \argmax_{\left(k,l,q \right)} \hspace{-1mm}\left(\hspace{-1mm} \left|p_{\left(k,l,q \right)}^{\left(\nu\right) }\right|^2 \hspace{-0.35cm}\sum_{\left(m,n,p\right)\in\mathcal{L}}\hspace{-3mm}  \left|\varphi_{\left(k,l,q \right)}\hspace{-0mm}\left[m,n,p\right]\right|^2 w\hspace{-0mm}\left[m,n,p\right]\hspace{-1mm}\right).
\end{equation}
The selected basis function is the one that minimizes the distance between the residual and the corresponding projection onto the basis function, weighted according to $w\left[m,n,p\right]$.

After having selected the basis function to be added, its weight has to be calculated. Since the basis functions cannot be mutually orthogonal as the scalar products can only be evaluated over $\mathcal{A}\cup\mathcal{R}$ where the weighting function $w\left[m,n,p\right]$ is unequal to zero \cite{Seiler2007}, the projection coefficient $p_{\left(k,l,q \right)}^{\left(\nu\right)}$ cannot be used directly as estimate. Instead, the fast orthogonality deficiency compensation proposed in \cite{Seiler2008} for 2D signals is adopted and
\begin{equation}
\hat{c}_{\left(u,v,z \right)} = \gamma \hat{p}_{\left(u,v,z \right)}  
\end{equation}
is used for obtaining an estimate for the expansion coefficient. In this context, $\gamma$ is fixed and from the range between zero and one. After basis function selection and weight estimation, the model and the residual of the current iteration are updated by
\begin{eqnarray}
\hspace{-3mm}g^{\left(\nu\right)}\hspace{-0.5mm}\left[m,n,p\right] \hspace{-3mm}&=&\hspace{-3mm} g^{\left(\nu-1\right)}\hspace{-0.5mm}\left[m,n,p\right] \hspace{-1mm}+\hspace{-1mm} \hat{c}_{\left(u,v,z \right)}\varphi_{\left(u,v,z \right)}\left[m,n,p\right]\\
\hspace{-3mm}r^{\left(\nu\right)}\hspace{-0.5mm}\left[m,n,p\right] \hspace{-3mm}&=&\hspace{-3mm} r^{\left(\nu-1\right)}\hspace{-0.5mm}\left[m,n,p\right] \hspace{-1mm}-\hspace{-1mm} \hat{c}_{\left(u,v,z \right)} \varphi_{\left(u,v,z \right)} \left[m,n,p\right].
\end{eqnarray}
The steps of selecting a basis function, estimating its weight and updating the model and residual are repeated until the predefined maximum number of iterations is reached. Finally, the real-valued component of the model in volume $\mathcal{B}_i$ is used as reconstruction for the unknown samples inside the currently considered cube. After replacing the unknown samples, the reconstruction proceeds to the next cube.

Unfortunately, the extrapolation process as outlined above is computationally very demanding. However, as shown in \cite{Meisinger2007, Seiler2010c} carrying out the model generation in the frequency domain can speed up the extrapolation dramatically. In \cite{Seiler2010c} it was also shown for the two-dimensional case that the real-valued model generation as it is used in \cite{Meisinger2007} is significantly slower than the proposed complex-valued model generation. Using the above presented complex-valued model generation and transferring the properties from \cite{Seiler2010c} to the three-dimensional case, the basis function selection and the expansion coefficient estimation can be expressed in the frequency domain by
\begin{equation}
\label{eq:bf_selection_freqdomain}
\left(u,v,z\right) =  \argmax_{\left(k,l,q\right)} \left|R_w^{\left(\nu-1\right)}\left[k, l,q\right]\right|^2
\end{equation}
and
\begin{equation}
\hat{c}_{\left(u,v,z \right)} = \gamma\frac{R_w^{\left(\nu-1\right)}\left[u,v,z\right]}{W\left[0,0,0\right]}
\end{equation}
with $W\left[k,l,q\right]$ being the DFT transformed weighting function $w\left[m,n,p\right]$ and $R_w^{\left(\nu-1\right)}\left[k,l,q\right]$ the transformed weighted residual
\begin{equation}
r_w^{\left(\nu-1\right)}\left[m,n,p\right] =  r^{\left(\nu-1\right)}\left[m,n,p\right]\cdot w\left[m,n,p\right].
\end{equation} 
With $G^{\left(\nu\right)}\left[k,l,q\right]$ as the DFT of $g^{\left(\nu\right)}\left[m,n,p\right]$, the update can be carried out in the Fourier domain, as well, by 
\begin{equation}
\label{eq:model_update_freqdomain}
G^{\left(\nu\right)}\left[u,v,z\right] = G^{\left(\nu-1\right)}\left[u,v,z\right] + MNP\hat{c}_{\left(u,v,z \right)}
\end{equation}
and
\begin{equation}
R_w^{\left(\nu\right)} \hspace{-1mm}\left[k,l,q\right] = R_w^{\left(\nu-1\right)} \hspace{-1mm}\left[k,l,q\right] - \hat{c}_{\left(u,v,z \right)} \hspace{-0.5mm}W \left[k\hspace{-1mm}-\hspace{-1mm}u, l\hspace{-1mm}-\hspace{-1mm}v, q\hspace{-1mm}-\hspace{-1mm}z \right]\hspace{-0.5mm},\hspace{-0.5mm} 
\end{equation}
for all $\left(k,l,q\right)$. Apparently, all operations for the model generation can be carried out in the Fourier domain and only transforms at the beginning and at the end of the modeling are required.


\section{Optimized Processing Order for Closing Three-Dimensional Holes} \label{sec:fse_reconstruction}

The generalized 3D-FSE as shown in the preceding section can be used very well for extrapolating an unknown signal within a cube, based on originally known or previously reconstructed samples. The question is how to process the different cubes in a volume if the region to be extrapolated is very large and spans multiple cubes. The obvious solution \cite{Meisinger2007} is to process the cubes in line-scan order, which is called 3D-FSE-LS in the following. However, by using a line-scan order, the reconstruction cannot account for the shape of the unknown region and it is not possible to exploit the available information from all directions effectively. An example for an extreme case would be if a large hole is located in the top left part at the beginning of the sequence. Then, no support samples would be available for modeling. In this case, the extrapolation of the first cube would be zero and this useless reconstruction would propagate to neighboring cubes due to the reuse of already reconstructed areas. Furthermore, a linescan order is not well suited for a parallel processing of cubes as neighboring cubes could not be processed independently without harming the reuse of already reconstructed cubes for supporting the further extrapolation. In order to cope with this, we propose an optimized processing order for closing three-dimensional holes using 3D-FSE. In the following, this will be called 3D-FSE-OPT.

The proposed 3D-FSE-OPT is inspired by an optimized reconstruction order for two-dimensional signals which has been shown in \cite{Seiler2011c}. For determining the actual order of the cubes to be processed, the feature of 3D-FSE to reuse already extrapolated areas is exploited. It has to be kept in mind that if already extrapolated cubes were not reused, a processing order would not be required, since every order would end in the same result. An important property of 3D-FSE is that the more samples are available in the support volume $\mathcal{A}$, the better typically gets the extrapolation quality. Using this, two criteria for determining the processing order can be defined:
\begin{enumerate}
	\item The more samples are available in support volume $\mathcal{A}$ surrounding error volume $\mathcal{B}_{i}$ the earlier it should to be extrapolated
	\item Directly adjacent cubes should not be processed at the same time since they then cannot support the extrapolation of each other
\end{enumerate}
Furthermore, it has to be considered that since the actual model generation already is computationally expensive, the determination of the processing order may not further increase the complexity. Hence, an easy criterion is required for determining the order. The proposed criterion to use is the number of not extrapolated neighboring cubes, depicted by $\mathcal{N}\left(\mathfrak{c}\right)$. For this, all cubes directly adjacent to the currently considered one are examined and if a neighboring cube contains samples which are unknown and have not been processed before, it is regarded as not extrapolated neighbor.

To start the extrapolation of a volume, it is divided into the cubes and for every cube the number of not extrapolated neighboring cubes has to be determined. For the initialization, $\mathcal{N}\left(\mathfrak{c}\right)$ is set to $0$, for all cubes. Then, the whole video sequence $v\left[x,y,t\right]$ is scanned cube by cube and if a cube $\mathfrak{c}$ contains samples to be extrapolated, $\mathcal{N}\left(\mathtt{neighbors}\left(\mathfrak{c}\right)\right)$ of all cubes adjacent to $\mathfrak{c}$ in spatial or in temporal direction is increased by $1$. In this context, the function $\mathtt{neighbors}\left(\mathfrak{c}\right)$ returns the indices of the cubes adjacent to the considered cube $\mathfrak{c}$. A special treatment is required for cubes lying at the outer margin of the sequence, since outside the video sequence no information is available. For a cube lying inside the volume, there are $26$ neighboring cubes available for supporting the extrapolation. However, for cubes located at the margin of the sequence, this number is smaller. Hence, for cubes located at the side,  $\mathcal{N}\left(\mathfrak{c}\right)$ is further increased by $9$, for cubes lying at the edge by $15$ and for the corner cubes by $19$. In order to mark all cubes without samples to be extrapolated, in a second scan $\mathcal{N}\left(\mathfrak{c}\right)$ is set to $-1$ for all cubes $\mathfrak{c}$ that do not contain any unknown samples.

After the initialization has been finished, the processing loop is started which lasts until all cubes are processed, or respectively, $\mathcal{N}\left(\mathfrak{c}\right)<0$ holds for all cubes. The loop starts with calculating the minimum non-negative number of not extrapolated neighbors
\begin{equation}
	\mathcal{N}_\mathrm{min} = \min_\mathfrak{c}\left(\max_\mathfrak{c}\left(0,\mathcal{N}\left(\mathfrak{c}\right)\right)\right).
\end{equation}
After this, all cubes are identified for which $\mathcal{N}\left(\mathfrak{c}\right)=\mathcal{N}_\mathrm{min}$ holds, which are the cubes that possess the most neighboring information, thus promising the highest extrapolation quality. These cubes are stored in list $\mathcal{S}$. For adding the cubes to the list, it has to be considered that a cube is only added to this list, if none of its direct neighbors is already on this list. In doing so, it can be assured, that never two directly neighboring cubes can be processed in the same run which is beneficial for two reasons. First, the reconstruction of one cube can be used for supporting the reconstruction of the other one. And second, all cubes on the list can be processed in parallel without getting a degradation due to the dependencies of the cubes. 

After the list $\mathcal{S}$ has been compiled it can be processed, that is to say, every cube on the list is extrapolated by the generalized \mbox{3D-FSE}. For every processed cube, $\mathcal{N}\left(\mathfrak{c}\right)$ is set to $-1$.  As the just extrapolated cubes can be used for supporting the extrapolation of their neighboring cubes and therewith improve the available neighboring information, $\mathcal{N}\left(\mathtt{neighbors}\left(\mathfrak{c}\right)\right), \forall \mathfrak{c}\in\mathcal{S}$ is decreased by $1$.

The steps of determining the cubes with best support and extrapolating them are repeated until all holes in the video sequence are filled, which means that $\mathcal{N}\left(\mathfrak{c}\right)<0, \forall \mathfrak{c}$. In order to provide a compact overview of the optimized processing order used in \mbox{3D-FSE-OPT}, Alg.\ \ref{algo:proc_order} lists the pseudo code of the algorithm. Using this algorithm, large holes can be closed from the outer margin to the center, therewith reducing the propagation of extrapolation errors and yielding a higher reconstruction quality as will be shown in the next section.

\begin{algorithm}[t]
	\caption{\emph{Hole filling by 3D-FSE with optimized processing order. Function $\mathtt{neighbors}\left(\mathfrak{c}\right)$ returns indices of cubes adjacent to cube $\mathfrak{c}$, $\mathcal{N}\left(\mathfrak{c}\right)$ holds the number of not extrapolated neighboring cubes for cube $\mathfrak{c}$.}}
	\fontsize{9}{7}\selectfont
	\label{algo:proc_order}
	\begin{algorithmic}
		\INPUT 3D input signal, divided into cubes
		\STATE /* Initialization */
		\FORALL {cubes $\mathfrak{c}$}
		\IF {cube $\mathfrak{c}$ has samples to extrapolate}
		\STATE $\mathcal{N}\left(\mathtt{neighbors}\left(\mathfrak{c}\right)\right) = \mathcal{N}\left(\mathtt{neighbors}\left(\mathfrak{c}\right)\right)+1$
		\ENDIF
		
		\IF {cube $\mathfrak{c}$ is located at corner of volume}
		\STATE $\mathcal{N}\left(\mathfrak{c}\right) = \mathcal{N}\left(\mathfrak{c}\right)+ 19$
		\ELSIF {cube $\mathfrak{c}$ is located at edge of volume}
		\STATE $\mathcal{N}\left(\mathfrak{c}\right) = \mathcal{N}\left(\mathfrak{c}\right)+ 15$
		\ELSIF {cube $\mathfrak{c}$ is located at side of volume}
		\STATE $\mathcal{N}\left(\mathfrak{c}\right) = \mathcal{N}\left(\mathfrak{c}\right)+ 9$
		\ENDIF
				
		\ENDFOR
		\FORALL {cubes $\mathfrak{c}$}
		\IF {cube $\mathfrak{c}$ has no samples to extrapolate}
		\STATE $\mathcal{N}\left(\mathfrak{c}\right) = -1$
		\ENDIF
		\ENDFOR
		\STATE /* Extrapolation */
		\WHILE {not all cubes finished}
		\STATE $\mathcal{N}_\mathrm{min} = \min\left(\max\left(0,\mathcal{N}\left(\mathfrak{c}\right)\right)\right)$
		\STATE $\mathcal{S} = \left\{\right\}$
		\FORALL {cubes $\mathfrak{c}$}
		\IF {$\mathcal{N}\left(\mathfrak{c}\right)== \mathcal{N}_\mathrm{min} \ \ \&\& \ \ \mathtt{neighbors}\left(\mathfrak{c}\right) \notin \mathcal{S}$}
		\STATE $\mathcal{S} = \mathcal{S} \cup \mathfrak{c}$
		\ENDIF
		\ENDFOR
		\STATE /* From here on: {\bfseries parallel} execution possible */
		\FORALL {cubes $\mathfrak{c}\in \mathcal{S}$}
		\STATE 3D-FSE for cube $\mathfrak{c}$
		\STATE Insert extrapolated samples in cube $\mathfrak{c}$
		\STATE $\mathcal{N}\left(\mathfrak{c}\right) = -1$
		\STATE $\mathcal{N}\left(\mathtt{neighbors}\left(\mathfrak{c}\right)\right) =\mathcal{N}\left(\mathtt{neighbors}\left(\mathfrak{c}\right)\right)- 1$
		\ENDFOR	
		\ENDWHILE
		\OUTPUT Extrapolated signal
	\end{algorithmic}
\end{algorithm}


\section{Simulations and Results}\label{sec:simulations} 

\begin{figure}
	\centering
	\psfrag{x}[c][c][0.75]{$x$}
	\psfrag{y}[c][c][0.75]{$y$}
	\psfrag{t}[c][c][0.75]{$t$}
	\psfrag{Linescan Order}[c][c]{Line scan order}
	\psfrag{Optimized Order}[c][c]{Optimized order}
	\includegraphics[width=0.49\textwidth]{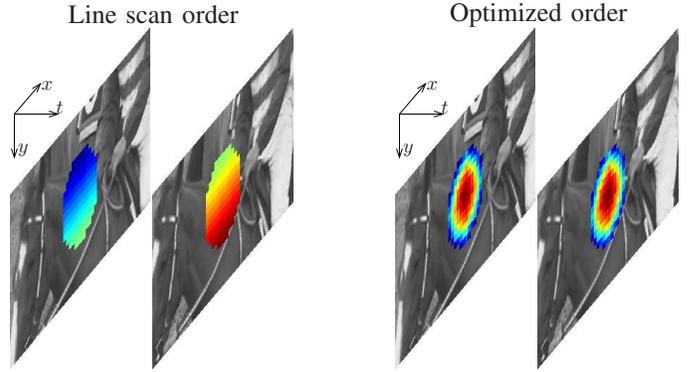}
	\caption{Example for order of cubes to be processed in two subsequent frames of a video sequence. Order of the processing is color coded, blue means early processing, red means late processing.}
	\label{fig:vis_order}
\end{figure}

In the following, the abilities and properties of the proposed optimized three-dimensional processing order in combination with the generalized 3D-FSE from Section \ref{sec:generalized_3D-FSE} will be analyzed. As stated above, the generalized 3D-FSE in combination with the proposed optimized processing order is depicted by 3D-FSE-OPT. In order to illustrate the effectiveness of the proposed processing order, \mbox{Fig.\ \ref{fig:vis_order}} shows the order of the cubes to be processed with respect to their location in the hole. For this, 3D-FSE-OPT is compared to \mbox{3D-FSE-LS}. Regarding the figure, it can be observed that \mbox{3D-FSE-LS} runs line by line and column by column through the hole, whereas 3D-FSE-OPT closes the volume from the outer margin to the center. In doing so, cubes in the center of the unknown area can benefit from supporting samples from all directions instead of only from cubes to the left, above, and in the past.

In order to assess the impact of the proposed optimized processing order on the reconstruction quality, the following test setup has been used. In addition to 3D-FSE-OPT and 3D-FSE-LS, two algorithms from literature are evaluated for comparison. This is the Video Inpainting (VI) algorithm from \cite{Newson2014} and the Three-dimensional Gap Filling (3D-GF) from \cite{Garcia2010, Wang2012}. The test setup consists of eight test sequences. For this, the four HEVC ClassC test sequences BasketballDrill, BQMall, PartyScene, RaceHorses of size $832\times 480$ pixels and the four HEVC ClassD test sequences BasketballPass, BlowingBubbles, BQSquare, RaceHorses of size $416\times 240$ pixels are considered. From all sequences, the first $48$ frames are tested. For 3D-FSE-OPT and 3D-FSE-LS the same parameters are used: the cube size is $4\times 4\times 4$, the border width is $14$ and an FFT of size $32\times 32\times 32$ is applied. The weighting function decays with $\hat{\rho}=0.7$, and $\gamma$ and $\delta$ are set to $0.5$.

\begin{table}
	\caption{Average PSNR (top in every row) and SSIM (bottom in every row) results for reconstruction of considered patterns. Best quality marked by bold font.}
	 \setlength{\tabcolsep}{1.5mm}
	\begin{tabular}{|l|c|c|c|c|}
		\hline  									& Diagonal bars & Lenses & Linear bars & Lifting  \\ 
		\hline VI \cite{Newson2014} 				& $25.15\punit{dB}$ & $25.23\punit{dB}$ & $25.04\punit{dB}$ & $26.43\punit{dB}$  \\
													& $0.9891$ & $\mathbf{0.9929}$ & $0.9895$ & $0.9817$\\													 
		\hline 3D-GF \cite{Garcia2010} 				& $21.87\punit{dB}$ & $21.65\punit{dB}$ & $21.07\punit{dB}$ & $23.21\punit{dB}$  \\ 
													& $0.9812$ & $0.9875$ & $0.9805$ & $0.9619$\\
		\hline 3D-FSE-LS 							& $25.98\punit{dB}$ & $25.34\punit{dB}$ & $25.72\punit{dB}$ & $27.27\punit{dB}$  \\ 
													& $0.9892$ & $0.9923$ & $0.9898$ & $0.9817$\\
		\hline 3D-FSE-OPT 							& $\mathbf{26.29\ dB}$ & $\mathbf{25.76\ dB}$ & $\mathbf{26.28\ dB}$ & $\mathbf{27.41\ dB}$  \\ 
													& $\mathbf{0.9896}$ & $0.9927$ & $\mathbf{0.9903}$ & $\mathbf{0.9821}$\\
		\hline 
	\end{tabular} 
	\label{tab:psnr_results}
\end{table}

\begin{figure*}
	\centering
	\psfrag{VI}[c][c][1]{\textcolor{white}{\bf VI}}
	\psfrag{3D-GF}[c][c][1]{\textcolor{white}{\bf 3D-GF}}
	\psfrag{3D-FSE-LS}[c][c][1]{\textcolor{white}{\bf 3D-FSE-LS}}
	\psfrag{3D-FSE-OPT}[c][c][1]{\textcolor{white}{\bf 3D-FSE-OPT}}
	\includegraphics[width=0.8\textwidth]{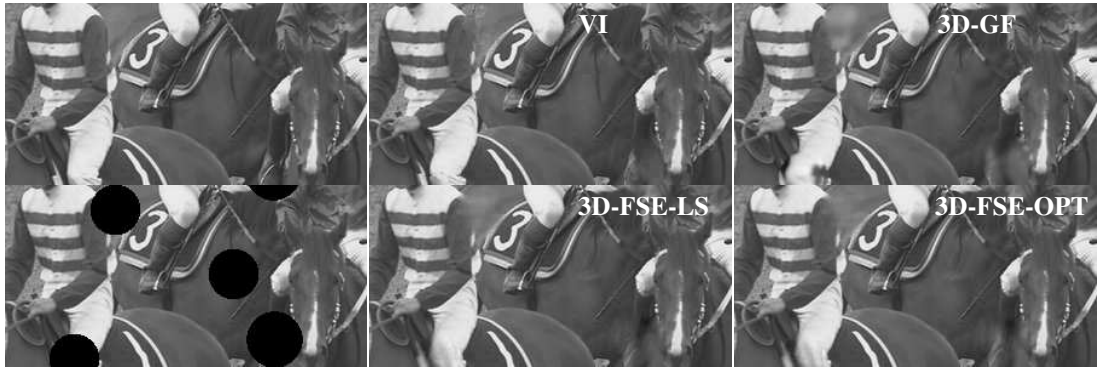}
	\caption{Detail of frame $15$ from sequence RaceHorses with error pattern \emph{Lenses}. First row: Original frame, reconstruction by VI \cite{Newson2014}, reconstruction by 3D-GF \cite{Garcia2010}. Second row: Frame with holes, reconstruction by 3D-FSE-LS, reconstruction by 3D-FSE-OPT.\vspace{-5mm}}
	\label{fig:visual_results}
\end{figure*}

Since the hole filling should operate as a generic approach and not be tailored to a specific application, four different patterns are examined. The patterns are very challenging, as they contain large consecutive unknown areas. The patterns of holes to be extrapolated are defined as follows:
\begin{itemize}
	\item \emph{Diagonal bars}: Eight bars of width $32\times 32$ samples running diagonally through the whole volume
	\item \emph{Lenses}: $30$ randomly placed lenses with radius $24$ samples in spatial direction and $4$ samples in temporal direction
	\item \emph{Linear bars}: $30$ randomly placed bars of spatial size $32\times 32$ samples and temporal size $12$ samples
	\item \emph{Lifting}: 3D error mask inspired by the output from compensated wavelet lifting \cite{Schnurrer2015}
\end{itemize}

Table \ref{tab:psnr_results} shows the average reconstruction quality in PSNR for the different patterns, evaluated over the holes. It can be observed that the optimized processing order yields significant gains over the linescan order of up to $0.56\punit{dB}$. At the same time, it can be seen that the algorithms used for comparison are outperformed by at least $0.53\punit{dB}$, in most of the cases by more than $1\punit{dB}$ PSNR and up to $5\punit{dB}$. The table also shows average SSIM results, evaluated over all frames of the sequences. Similar to the PSNR results, it can be observed that the optimized processing order achieves a very high quality and is superior to the other algorithms, except for one case where it is slightly outperformed by VI.

Aside from the objective evaluation, the visual quality of the reconstruction is of high importance, as well. To illustrate, that 3D-FSE-OPT can also yield a higher visual quality, especially in comparison to 3D-FSE-LS, Fig.\ \ref{fig:visual_results} shows frame $15$ of sequence RaceHorses with pattern \emph{Lenses}. Please note that the hole extends to neighboring frames as well but only one frame is shown here. Comparing the results for the different algorithms, it can be observed that the ability of 3D-FSE-OPT to close the holes from all directions leads to reduced distortions compared to 3D-FSE-LS and overall 3D-FSE-OPT yields the highest visual quality.


\section{Conclusion and Outlook} \label{sec:conclusion} 

In this paper an optimized processing order was introduced for filling arbitrarily shaped holes in video sequences by \mbox{3D-FSE}. By making the order of the processing adaptive to the shape of the area to be extrapolated, a higher reconstruction quality can be achieved, leading to visually noticeable gains of more than $0.56\punit{dB}$ PSNR compared to a fixed linescan processing. Using the proposed processing order, the generalized 3D-FSE also is superior to other reconstruction algorithms, and gains of several $\punit{dB}$ PSNR are possible. Aside from leading to a higher extrapolation quality, the proposed order inherently allows for a parallel processing of the cubes, making it suited more for implementation on state-of-the-art compute platforms. Future research aims at generalizing the optimized three-dimensional processing order and transferring it to alternative spatio-temporal extrapolation algorithms.




\end{document}